# Decoupling and tuning competing effects of different types of defects on flux creep in irradiated YBa$_2$Cu$_3$O$_{7-\delta}$ coated conductors

S. Eley[1], M. Leroux[2], M. W. Rupich[3], D. J. Miller[4], H. Sheng[4], P. M. Niraula[5], A. Kayani[5], U. Welp[2], W.-K. Kwok[2], L. Civale[1]

[1]Materials Physics and Applications Division (MPA) and Condensed Matter and Magnet Science (CMMS), Los Alamos National Laboratory, Los Alamos, NM 87545 USA
[2]Materials Science Division, Argonne National Laboratory, Argonne, IL 60439 USA
[3]American Superconductor Corporation, 64 Jackson Rd, Devens, MA 01434 USA
[4]Center for Nanoscale Materials, Argonne National Laboratory, Argonne, IL 60439 USA
[5]Department of Physics, Western Michigan University, Kalamazoo, MI 49008 USA

**Abstract**.

YBa$_2$Cu$_3$O$_{7-\delta}$ coated conductors (CCs) have achieved high critical current densities ($J_c$) that can be further increased through the introduction of additional defects using particle irradiation. However, these gains are accompanied by increases in the flux creep rate, a manifestation of competition between the different types of defects. Here, we study this competition to better understand how to design pinning landscapes that simultaneously increase $J_c$ and reduce creep. CCs grown by metal organic deposition show non-monotonic changes in the temperature-dependent creep rate, $S(T)$. Notably, in low fields, there is a conspicuous dip to low $S$ as the temperature ($T$) increases from ~20 K to ~65 K. Oxygen-, proton-, and Au-irradiation substantially increase $S$ in this temperature range. Focusing on an oxygen-irradiated CC, we investigate the contribution of different types of irradiation-induced defects to the flux creep rate. Specifically, we study $S(T)$ as we tune the relative density of point defects to larger defects by annealing both an as-grown and an irradiated CC in O$_2$ at temperatures $T_A$ = 250°C to 600°C. We observe a steady decrease in $S(T > 20$ K$)$ with increasing $T_A$, unveiling the role of pre-existing nanoparticle precipitates in creating the dip in $S(T)$ and point defects and clusters in increasing $S$ at intermediate temperatures.

## I. INTRODUCTION

YBa$_2$Cu$_3$O$_{7-\delta}$ (YBCO) is the most studied superconductor, and decades of enduring interest in engineering the vortex pinning landscape in this material have effectuated remarkable increases in the critical current density ($J_c$). Vortex motion, that limits the $J_c$, can be suppressed by introducing defects using a variety of techniques. Notably, particle irradiation can create vacancies and interstitials, larger random defects, or correlated disorder (e.g., amorphous tracks) while a variety of chemical processes can incorporate nanoparticles (NPs) or nanorods of non-superconducting secondary phases.[1] The performance (specifically, $J_c$) of YBCO single crystals, thin films, and coated conductors (which are rare-earth doped epitaxial YBCO films on flexible metal substrates) all benefit from these techniques. High quality as-grown YBCO single crystals have weak pinning and self-field $J_c$ of approximately 1-3 MA/cm$^2$ at 4 K; particle irradiation can dramatically increase these values to 20-30 MA/cm$^2$.[2] However, the $J_c$ of as-grown epitaxial YBCO films and coated conductors (CCs) can be even higher ($J_c$>50 MA/cm$^2$)[3] indicating that the complex pinning landscapes that naturally form in these systems are already quite effective. In fact, such samples exhibit the highest attained $J_c$ and $J_c/J_0$ of any technologically relevant superconductor; here, $J_o$ is the depairing current density, [4] considered the theoretical upper limit for $J_c$. However, irradiation with protons or oxygen ions can further double $J_c$ at high fields. A great advantage of the latter technique is that it is fast and inexpensive enough to make it potentially viable for implementation in continuous CC manufacturing processes. [5]

Further improvements in the performance of CCs requires a deeper understanding of the response of vortex matter to their intricate mixed pinning landscapes, in which the contributions of the different type of defects are clearly non-additive, being synergistic in some cases and competitive in others. Thus far, the main parameter used to quantify the effectiveness of various pinning landscapes has been the measured $J_c$. We note, however, that flux creep (the thermally activated process that allows vortices to depin even for current densities below $J_c$) substantially reduces the *effective* (measured) $J_c$ in high-temperature superconductors (HTS). This is possible because flux creep makes the measured $J_c$ dependent on the measurement time scale, or equivalently, on the specified electric field of the measurement. In fact, flux creep essentially halves $J_c$ in YBCO at $0.5T_c$ and 1T from the value it would have had in the absence of flux creep. Moreover, the increases in $J_c$ due to irradiation are frequently accompanied by increases in creep. [5-7] Thus, searching for methods to reduce creep and understanding the connection between flux creep and $J_c$ must be considered an important aspect of pinning engineering. Furthermore, as we will show below, creep studies can provide unique additional information about the interplay of the various pinning mechanisms.

The normalized creep rates ($S = d \ln M / d \ln t$, where $M$ is the magnetization and $t$ is time) of all YBCO samples, including as-grown, irradiated and chemically engineered single crystals, films and CCs, are very similar, even though their critical currents can differ by orders of magnitude. [2,5-11] Some examples are shown in Figure 1. At low fields, $S$ typically increases approximately linearly with $T$ up to ~20 K. For $T$ between 20 K and 50 K, $S(T)$ shows a plateau that is often attributed to a vortex glass state. At even higher temperatures, $S$ increases rapidly as the system crosses over into the plastic regime. An exception to this behavior is observed in CCs grown using metal organic deposition (MOD), [10,12] which show lower creep rates for a broad range of temperatures, including a characteristic dip at intermediate temperatures, as seen in Figure 1. The source of this dip is not fully understood, but a correlation between the existence of the dip and NPs is incontrovertible. In fact, measurements of MOD-grown $Y_{0.77}Gd_{0.23}Ba_2Cu_3O_y$ coated conductors with a low density of large $Y_2Cu_2O_5$ precipitates display a dip in $S(T)$ that deepens upon the addition of a high density of $BaZrO_3$ nanoparticles. [10,12] Investigating vortex dynamics in MOD CCs is relevant not only because their low creep rates are advantageous for applications, but also because understanding why these systems have the lowest $S$ values in any YBCO material may lead to the discovery of more general strategies to reduce creep.

In this study, we explore changes in $S(T, H)$ and $J_c(T, H)$ as we tune the ratio of smaller (point to few-nm-sized) defects to larger nanoparticles in an as-grown and an oxygen-irradiated MOD CC by annealing both samples in oxygen at temperatures ranging from 250°C to 600°C. The main results can be summarized as follows. First, we provide evidence that the dip to low $S$ in the as-grown CC (see Figure 1) is due to strong pinning from individual NPs. Second, the introduction of a dispersion of smaller defects from irradiation drastically increases $S$ (reduces the dip depth), indicative of competing effects. Third, the dip is progressively recovered as point defects are preferentially removed during the annealing process. Our results indicate that additional point defects are the predominant cause of the increase in $S$ after irradiation and that the reduction in $S$ produced by NPs can be counteracted by the presence of a high enough density of additional defects.

Decoupling pinning effects based on the defect size, distribution of sizes, and presence of other defects is a burgeoning focus of research in this field. Numerical simulations [13] based on the time-dependent Ginzburg-Landau (TDGL) model have recently been successful in reproducing experimental results for the critical current in systems with correlated defects [14] and nanoparticles of various sizes and densities. [15,16] However, there is no similarly accurate modeling of creep in such systems, and experimental studies are complicated by the difficulty in controlling the size of dispersed defects. [17]

We use annealing as a tool to systematically and selectively mobilize the smaller defects, [18] which diffuse faster than larger ones. Annealing effects in YBCO have been extensively studied. [19-23] Recovery of the lattice by defect annihilation occurs if these defects can recombine within an effective diffusion range $X = \sqrt{D(T)t_a}$, where $D(T)$ is the temperature-dependent diffusion constant and $t_a$ the annealing time. Point defects will become mobile in both the as-grown and oxygen-irradiated samples during annealing. Theoretical Frenkel-pair annihilation curves [19] calculated from the Ullmaier-Schilling model predict recombination of oxygen Frenkel pairs between 100°C and 200°C, and copper Frenkel pairs between 450°C and 600°C. We do not anneal at high enough temperatures to affect Frenkel pairs of heavier ions (e.g., Ba, Y) [19] nor activate the larger, more thermodynamically stable rare-earth-oxide nanoparticles.

This simple scenario is complicated by the fact that, at temperatures higher than 450°C, the film will slightly deoxygenate. [24] Deoxygenation changes $T_c$ due to changes in the density of current carriers; reduces the effective bulk pinning [25,26] due to a decrease in $\xi_c$, hence an increase in anisotropy $\gamma = \xi_{ab}/\xi_c$; decreases the thermal activation energy $U(T,H)$ for vortex motion, [27,28] and increases the number of oxygen vacancies (point defects). [25] To decouple these concomitant effects above 450°C, we partially reoxygenate each sample by annealing in oxygen at 300°C for 3-5 hours after the last annealing step. (Full reoxygenation would require annealing at 300°C for a few days). [24] This annealing procedure will prove a useful tool to decouple the effects of nanoparticles, clusters, and point defects on creep.

This work is organized as follows. In section II, we describe the architecture and microstructure of the as-grown and irradiated samples, the annealing sequences, and magnetization measurement procedures. In section III.A, we show how irradiation and annealing affect $J_c$ and $T_c$. Sections III.B and III.C, the central focus of this paper, detail the changes in the creep rate upon introduction of irradiation-induced and subsequent removal of point defects. We consider if these changes could be instigated by a change in the pinning regime; specifically, we analyze whether irradiation introduces enough defects to create a crossover from a regime of individual, strong pinning to weak collective pinning.

We then paint a phenomenological picture of how the addition of irradiation-induced defects could lower the activation barrier, resulting in increased creep. Finally, in section IV we present our conclusions.

## II. SAMPLES AND EXPERIMENTAL PROCEDURES

The samples studied here are CCs cut from commercial tape that was manufactured at American Superconductor Corporation (AMSC) and patterned into 1.5 mm x 1.5 mm squares using photolithography and Ar-ion milling. Each consists of a 1.2-µm-thick $YBa_2Cu_3O_{7-\delta}$ film, doped with $Dy_2O_3$, deposited on a 75-nm-thick $Y_2O_3$, YSZ, and $CeO_2$ buffer layer on a Ni(5at%)W substrate, and capped with a 1-µm Ag layer. The YBCO layer is grown using the MOD process that is described elsewhere [29,30] and annealed post-growth at 400°C. AMSC has optimized the microstructure of the as-grown sample for commercial use at low temperatures and high fields. Planar defects such as stacking faults and twin boundaries, dislocations, point defects (vacancies, interstitials, atomic-scale impurities) and rare earth oxide nanoparticles define the pinning landscape of the as-grown films. [31] The nanoparticles, which are significantly larger (20-50 nm diameter) than point defects, are precipitates of secondary phases ($Dy_2O_3$ additions) and are visible in the diffraction contrast transmission electron micrographs (TEM) in Figure 2(a). We estimate the density of nanoparticle defects to be about $2\times10^{14}$ cm$^{-3}$ occupying a volume fraction of ~1.2%. One sample was irradiated with $^{16}O^{3+}$ ions at a dose of $5\times10^{13}$ O/cm$^2$ and energy of 6.0 MeV (at the sample surface) using a tandem Van de Graff accelerator at Western Michigan University. SRIM-TRIM simulations predict that this energy is high enough that the Bragg peak lies in the substrate and the distribution of defects is defined by the tail of the Bragg peak, increasing from small values near the Ag-YBCO interface to ~ $10^{21}$ vacancies/cm$^3$ at the YBCO-buffer layer interface. [5]

Diffraction contrast TEM studies on oxygen-ion irradiated MOD CCs revealed anisotropic defect clusters of approximately 5 nm in size. [5] Similar defects were also observed in CCs irradiated with 4-MeV protons. [7] From the TRIM predictions and the TEM observations, we estimate the density of these clusters to be roughly $2\times10^{16}$ cm$^{-3}$, therefore occupying a volume fraction of 1.3%. Extensive microstructural studies on YBCO single crystals [32,33] showed that irradiation generally produces a mixture of point defects in the form of Frenkel pairs of interstitials and vacancies, clusters in the form of dislocation loops or platelets, and collision cascades. These studies also revealed that the irradiation-induced defect structure depends on the pre-existing microstructure. We have not identified the exact morphology of the clusters in our oxygen-ion irradiated CCs, but imaging suggests they are agglomerations of interstitials. Cascades and clusters are visible in TEM whereas point defects are not. Hence, irradiation introduces multiple changes in the sample microstructure such that the contribution of different types of irradiation-induced defects to changes in $S$ and $J_c$ is not immediately evident.

An as-grown reference sample and the oxygen-irradiated sample were both subjected to a progression of annealing cycles. Using a tube furnace with 1 atm of flowing oxygen gas, each sample was annealed 9 times over the course of the study at progressively higher temperatures, $T_A$, ranging from 250°C to 600°C, in increments of 50°C. A final annealing step was performed at 300°C for 3 hours to reoxygenate the sample. Each annealing cycle involved heating the sample at a rate of 20°C/min, maintaining the maximum temperature for 30 minutes, then cooling at approximately 25°C/min down to 150°C, after which the sample was allowed to cool naturally in an oxygen environment.

TEM images (e.g., Figure 2(c)) of the oxygen-irradiated sample post-annealing reveal the residual presence of many irradiation-induced clusters, evincing the robustness of these defects to the annealing sequence. This suggests that, as previously described, the predominant effect of the annealing steps is to progressively reduce the density of point defects. Such selectivity allows us to dissociate pinning dynamics caused by the clusters from that caused by point defects.

Magnetization studies were performed prior to annealing and after each annealing step using a Quantum Design SQUID magnetometer to determine the critical temperature, $T_c$, and characterize the temperature and field dependence of $J_c$ and $S$. The magnetization ($M$) was obtained by dividing the measured magnetic moment ($m$) by the sample volume $V = \delta w l$, where $\delta$ = 1.2 µm is the film thickness, and $w = l$ = 1.5 mm specifies sample width and length. For all measurements, the magnetic field was applied parallel to the c-axis (perpendicular to the film plane). $T_c$ was determined from magnetization as a function of temperature at $H$ = 2 G. The critical current was calculated using the Bean critical-state model,[34,35] $J_c = 20\Delta M / [w(1-\frac{w}{3l})]$, where $\Delta M$ is the difference between the upper and lower branches of isothermal magnetization loops $M(H)$. Creep data were taken by using standard methods [36] for measuring the time-dependent magnetization $M(t)$. First, the critical state was established by sweeping the field $\Delta H > 4H^*$, where $H^*$ is the minimum field at which magnetic flux will fully penetrate the sample. Magnetization is then recorded every ~15 s; a brief measurement in the lower branch is collected to determine the background (arising predominantly from the NiW substrate), then the upper branch is measured as a function of time ($t$) for an hour in fixed fields of 0.3 T, 1.0 T, or 5.0 T. After subtracting the background and adjusting the time to account for the difference between the initial application of the field and the first measurement (maximize correlation coefficient), $S = d \ln J_c / d \ln t$ is extracted from the slope of a linear fit to ln $M$-ln $t$.

### III. RESULTS

#### A. Critical Current

The evolution of $T_c$ and of $J_c$ at $T$ = 5 K, normalized to the pre-annealed as-grown value, as a function of $T_A$ for both samples is shown in Figures 3(a) and 3(b), respectively. Figure 3(b) shows how irradiation increases $J_c$, while annealing can reverse this enhancement by removing point defects. The non-monotonic behavior of both $T_c$ and $J_c$ is qualitatively indicative of the different stages in which the defects anneal, as labeled. Region I captures the pre-annealing condition of the samples. Consistent with results we reported recently,[5] at 5 K, oxygen irradiation increases $J_c$ by a factor of more than 2.5 at 5 T while it increases $J_c$ by a factor of 1.5 at 1 T and has little effect at 0.3 T. The increased effectiveness at higher fields indicates that the irradiation introduces a large density of small pinning sites.[5,7] Oxygen irradiation reduces $T_c$, an effect also seen in proton irradiated samples, presumably because the irradiation-induced point defects behave as nonmagnetic scattering centers.[37] For annealing up to 300°C (region II), diffusion and rearrangement of mobile oxygen anions [19,20] cause a decrease in disorder, increasing $T_c$ and decreasing $J_c$ in the irradiated sample. In Region III, $J_c$ increases slightly in the irradiated sample due to decreased anisotropy resulting in increased effective bulk pinning, while $T_c$ remains almost constant. As expected, we observe minimal differences in $T_c$ and $J_c$ in the as-grown sample in regions I, II, and III because it previously underwent post-growth annealing at 400°C at AMSC. Further heating above 450°C (region IV) promotes removal of more point defects, including copper Frenkel pairs, causing a reduction in $J_c$ in the irradiated sample. The drop in $J_c$ in the oxygen-irradiated sample is consistent with a reduction in the number of additional defects that do not exist in the as-grown sample. The $J_c$ values for the oxygen-irradiated sample are similar after the 500°C and 550°C annealing steps, and for 1 T and 5 T they correspond to about 40% retention of the irradiation-induced enhancement. This suggests that most of the point defects have been annealed out and the remaining extra pinning arises from the clusters. A second order effect in this region is the deoxygenation. Deoxygenation has been observed in irradiated YBCO single crystals at $T_A > 450$°C (in flowing $O_2$) [24]. The temperature at which and extent to which deoxygenation occurs will depend on the annealing time, temperature, and oxygen pressure, and the onset is associated with a broadening of the transition width [24], which we first observe at 500°C. Above 550°C (region V), the pronounced drop in $J_c$ at all fields, which reverses most of the remaining $J_c$ enhancement gained from irradiation, suggests that we are now substantially deoxygenating the sample. Reoxygenation creates a 10% increase in $J_c$(1 T) at $T$ = 20 K, 35 K, and 40 K (not shown). Similarly, the changes in $T_c$ in the last 3 annealing stages may be related to changes in oxygen stoichiometry caused by deoxygenation then reoxygenation of the samples.

 The irradiation-induced enhancement in $J_c$ has a clear temperature and field dependence. Figure 4(a) shows $J_c(T)$ in the as-grown sample, the oxygen-irradiated sample, and after annealing the latter at 550°C. Figure 4(b) highlights the temperature and field dependence of the enhancement factor, $J_c/J_c^{As-grown}$. One salient feature is that the enhancement at 5 T extends over the full temperature range. At 1 T, the range of enhancement is limited to $T < 60$ K, and at 0.3 T, the enhancement is minimal at low $T$, and irradiation is detrimental to $J_c$ for $T > 20$ K. The inset of Figure 4 shows the line in the H-T plane separating the regions in which oxygen irradiation increases and decreases $J_c$. Clearly, irradiation is beneficial over the vast majority of the phase diagram, in particular in the region of interest for rotating machines[27,28] ($H \approx$ 2-5 T and $T \approx$ 20-40 K). For $H$ = 5 T, the enhancement in $J_c$ extends over the entire vortex solid phase, up to the irreversibility line, implying that oxygen irradiation would be beneficial for wires constituting very high field magnets.

 Now, the reduction of $J_c$ at low $H$ and high $T$ is provocative, implying a competition between the pre-existing and the added defects. As shown in the inset to Figure 4, this covers a very small portion of the $H$-$T$ plane; however, a detailed understanding of the origin of this behavior is important for several reasons. Wires for transformers, fault current limiters, and cables operate in this regime ($H \approx$ 0.1-0.5 T), therefore, would not benefit from oxygen irradiation. Secondly, it may allow us to devise a strategy to reduce this competition. Competition among different types of pinning centers is ubiquitous, possibly also occurring in other regions of the *H-T* plane, yet obscured by an overall positive effect. As shown in Figure 4, annealing the irradiated sample at $T_A$=550°C drives $J_c(T)$ towards its as-grown value at all fields and a wide range of temperatures. Particularly interesting is the case of $\mu_0 H = 0.3$ T and 1 T at high *T*, where the elimination of defects by annealing results in an increase in $J_c$. This confirms the existence of competing effects between the preexisting defects in the as-grown film and the irradiation-induced point defects in this region of the H-T plane.

We further investigate the effects of annealing on $J_c$ by measuring the field dependence at *T* = 5 K and 40 K. Figure 5(a) shows $J_c$ vs *H* at *T* = 5 K for the as-grown sample, the oxygen-irradiated sample, and the latter after each annealing step. We find the usual *H*-independent $J_c$ regime at low *H*, as well as the frequently observed power law regime in which $J_c \propto H^{-\alpha}$ for *H* > 1 T. Focusing first on comparing the pre-annealing data for each sample, we see that the extracted power law exponent α is significantly reduced by irradiation from ~0.7 to below 0.4, consistent with previous reports on CCs irradiated with oxygen ions,[5] protons,[6,7] and silver ions.[38] Notably, $\alpha \approx 0.7$ is close to the predicted value, 0.625, for pinning by large nanoparticles.[39,40] Changes in $\alpha$ after irradiation suggest that the additional pinning arises mainly from the smaller defects. This is consistent with results from a study of proton-irradiated CCs, in which α monotonically decreases with increasing dose.[6] We now consider the effects of annealing. The evolution of $\alpha$ with $T_A$ at *T* = 5 K and *T* = 40 K is plotted in Figure 5(c). In the case of the as-grown sample, the exponent α does not change significantly, indicating no change in the dominant pinning mechanism. As annealing reduces the number of point defects, this indicates that point defects are not the dominant pinning mechanism. In the irradiated sample, $\alpha$ monotonically increases with $T_A$, demonstrating the progressively decreasing contribution of point defects to pinning. After the 600°C anneal, $\alpha = 0.62$, as expected for pinning from large nanoparticles. The fact that $\alpha$ does not fully recover the pre-irradiation value after the highest $T_A$ means that the remaining clusters and cascades are affecting it. Clusters appear to have similar pinning properties to rare-earth-oxide NPs, such that the combination of NPs and clusters results in a strongly NP dominated landscape, with $\alpha$ coinciding with the predicted value.

Pinning results from spatial variations of the Ginzburg-Landau order parameter, capturing disorder in either $T_c$ or in the charge carrier mean free path $\ell$ near the defects.[41] For single vortex pinning, $J_c$ is predicted to depend on the temperature as $J_c(T/T_c) \propto [1 - (T/T_c)^2]^n$, with *n* = 2.5 for $\delta\ell$ pinning expected in the case of a completely point defect dominated landscape, and *n* = 1.2 for $\delta T_c$ pinning, expected for nanoparticles.[42] Figure 5(b) shows $\log(J_c)$ vs. $\log[1 - (T/T_c)^2]$ at $\mu_0 H = 1$ T for the oxygen-irradiated sample after each annealing step. From the slopes of the linear dependences at high *T* ($T/T_c \geq 0.55$), we can extract *n* as shown in Figure 5(c), which also includes analogous data for the as-grown sample. Given the complex, mixed pinning landscape in our CCs and the effects of creep on *n*, we do not expect them to exhibit the exact *n* values predicted for $\delta\ell$ or $\delta T_c$ pinning. It is clear, however, that irradiation increases *n*, consistent with a shift from a NP-dominated pinning landscape towards one dominated by point defects. As is the case with α, *n* remains unchanged for the as-grown sample, indicating no change in the dominant pinning mechanism. The irradiated sample shows a clear progressive recovery towards a NP dominated landscape; after annealing at 500°C, when most of the point defects have been removed but the clusters are still present, *n* has returned to pre-irradiation values.

## B. **Flux creep in the as-grown and irradiated samples**

To better understand the competition between defects, we focus on the temperature and field dependence of flux creep, a dynamic effect caused by the thermally activated motion of vortices. From creep measurements, we can extract information about the pinning energies of defects. We can also characterize the nature of pinning, specifically whether vortex segments are predominately pinned by large, individual defects or by the collective action of many smaller, weaker defects.

Figure 6 shows $S(T)$ at $\mu_0 H$ = (a) 0.3 T, (b) 1 T, and (c) 5 T for the as-grown sample, the irradiated sample, and the latter after two of the annealing steps. In all cases, $S(T)$ increases with $T$ up to 20-25 K. For $\mu_0 H$ = 0.3 T and 1 T, $S(T)$ is non-monotonic, reaching a local maximum at $T \approx$ 25 K and 20 K, respectively, then dipping down to a broad local minimum at $T \approx$ 45 K and 35 K, then increasing again as $T$ is increased further. For $\mu_0 H$ = 5 T, $S(T)$ is monotonic, with a short plateau between $T \approx$ 15 K and 25 K. Similar curves have been reported and discussed in the past.[5,7,10,12] In this section, we will analyze the results for the as-grown and irradiated samples prior to annealing, and in the next section, we will discuss the effects of annealing these samples.

At low $T$ (i.e., $T$ < 20 K), $S(T)$ for both samples is similar. As seen in Figures 6(a-c), this is true at all fields. This is a universal result for YBCO. For example, it is also the case for all samples in Figure 1, including the single crystal, which has much lower $J_c$, and can also be observed in published data on a variety of YBCO crystals and films.[7,10,11] According to the Anderson-Kim (A.K.) description of flux creep, at low temperatures $S(T) \approx k_B T/U_P(T)$ where $U_P$ is the pinning energy. Ignoring the $T$ dependence of $U_P$, the approximately linear increase in $S$ with $T$ up to the local maximum is qualitatively in agreement with A.K., with one glaring exception: the curves extrapolate to finite $S(T = 0) > 0$, indicative of quantum creep.

The order of magnitude of $U_P$ can be estimated by considering individual vortex pinning by a defect of the size of a coherence volume $V_1 \sim (4\pi/3)\xi_{ab}^2 \xi_c = (4\pi/3)\xi_{ab}^3/\gamma$, thus $U_{P,1} \sim (H_c^2/8\pi)V_1 \sim \varepsilon_0 \xi_{ab}/3\gamma$, where $H_c$ is the thermodynamic critical field, $\varepsilon_0 = \Phi_0^2/(4\pi\lambda_{ab})^2$ is the vortex energy scale, $\gamma \approx 5$, and $\Phi_o$ is the flux quantum. For YBCO at low temperatures $U_{P,1} \sim$100 K, while the slope d$S$/d$T$ in Figures 6(a) implies that $U_P \sim$700 K. Pinning energies larger than $U_{P,1}$ can be achieved through individual strong pinning by defects larger than $V_1$, however the largest achievable $U_P$ is ultimately limited by the elastic properties of the vortex matter. For instance, a vortex can depin from an arbitrarily long columnar defect by producing a "half loop" of length $\ell_{hl} \sim \xi_c(J_0/J)$ along the columnar defect. In the early stages of the relaxation process, at low $T$ (where $J$ is close to $J_c$), $\ell_{hl}$ is only $\sim J_0/J_c$ times $\xi_c$, resulting in pinning energies $U_{P,hl} \sim \varepsilon_0(\xi_{ab}/\gamma)(J_0/J_c)$.[41] For NPs, Koshelev and Kolton [43] have shown that with increasing $J$, the tips of the pinned vortex slide along the surface of the NP until they meet near the NP equator and reconnect, leading to depinning of the vortex. This process determines $J_c$ and implies that the pinning force is limited by the vortex line tension to $F_P \lesssim (2\varepsilon_0/\gamma)\ln(R/\xi_c)$. As $J$ decays during the relaxation process, the two tips of the vortex progressively separate. In the early stages, when $J$ is close to $J_c$, the separation between the tips is small and any thermal fluctuation that creates a short vortex segment that connects the tips is enough to depin the vortex. The activation energy associated with this depinning process is small, regardless of the NP radius $R$. At low $T$ the creep rates widely observed in YBCO appear to be limited by the Ginzburg number (which measures the strength of thermal fluctuations) [41] and the elastic vortex properties.

As $T$ increases, $U_{P,1}(T)$ decreases and $S(T)$ should increase faster than linearly with $T$, in contrast to the observation. Thus, some mechanism must account for the increase in the effective pinning energy with increasing $T$. It has long been recognized that this may occur in the case of collective creep, in which the length of the depinned vortex segment or volume of the depinned bundle increases as $J_c$ decreases (i.e., $T$ increases), resulting in an effective activation energy that increases with $T$. However, a similar effect may occur even without collective effects, [44] for instance in strong single vortex pinning by columnar defects where $\ell_{hl}$ increases with decreasing $J$ (i.e., with increasing $T$). Analogously, for NPs the pinning potential strongly increases with decreasing $J$ as the two tips of the vortices recede from each other and the segment needed to reconnect them becomes longer. In the limit of zero current, the pinning potential corresponding to core pinning by the NPs is reached and $U_P \sim 2\varepsilon_0 R \ln(R/\xi_{ab})$. [43]

The most striking difference between $S(T)$ for the as-grown and the irradiated sample is the depth of the dip in $S$ at intermediate temperatures (between 20 K and 60 K). As previously mentioned, in the as-grown sample we measure a deep dip in $S$ and, in the irradiated one, a significantly shallower dip evocative of the plateau frequently observed in YBCO and some Fe-based superconductors. [45–47] Previous studies have correlated the dip with the presence of NPs, thus we expect a high sensitivity to field as the ratio of vortices to that of available pinning sites changes. Consistently, Figures 6(a), 6(b), and 6(c) show that the dip is largest for $\mu_0 H$ =0.3 T, smaller for 1 T, and nonexistent for 5 T. The intervortex distance $a_o = [2\Phi_o/(\sqrt{3}B)]^{1/2}$, where $B$ the magnetic field, is 89 nm at 0.3 T, 49 nm at 1.0 T, and 22 nm at 5 T, compared to an estimated average distance of ~167 nm between NPs in both samples. We see a similar sensitivity to the number of available pinning sites from the dependence of $S(T)$ on irradiation dose, in which the dip in $S(T)$ at a fixed field systematically flattens with increasing O-ion dose. [5]

In the following discussion, we examine whether pinning in our samples is due to the independent action of strong, individual defects or the collective action of many weaker defects. In the as-grown sample, as previously discussed, the critical current data indicate that at low $H$ the vortex pinning and dynamics are dominated by the NPs. For simplicity, let's assume that $J_c$ comes only from the NPs, ignoring other contributions. At 0.3 T, the Larkin length $L_c \approx (\xi_{ab}/\gamma)(J_o/J_c)^{1/2}$, which sets the length scale for collective pinning, [41,48] is ~ 0.9 nm and 1.8 nm at 5 K and 45 K, respectively. (Note that by using the measured $J_c$, which is affected by creep, we overestimate $L_c$). This is orders of magnitude smaller than the average distance between NPs, confirming that pinning is individual rather than Larkin-Ovchinnikov-like (collective). The non-monotonic $S(T)$ observed at low $H$ can occur in the regime of strong single vortex pinning by individual defects if the activation barrier $U(J)$ of the NPs increases steeply as the current density $J$ approaches zero, as shown in Figure 7(a). This is indeed the case for large NPs as discussed above. In the as-grown sample ($R_{NP} \sim 10$ to 25 nm), at $T$ = 45 K this gives $U_P(J \approx 0) \approx 17000$ K $-$ 70000 K. In our creep measurements, $J$ at 45 K is still quite large ($\sim 10$ MA/cm$^2$), thus we expect $U_P$ to be between these values and that found at low temperatures $U_P(J \sim J_c) \approx 700$ K. In fact, as shown in Figure 6(a), the minimum $S(T = 45$ K$, H = 0.3$ T$) \approx 0.016$ corresponds to an effective activation energy $T/S \approx 2800$ K.

As previously mentioned, oxygen irradiation introduces a dispersion of clusters and cascades of average radius $R_{cl} \sim 2.5$ nm, spaced 38 nm apart on average, and an unknown density of smaller defects that are not resolved with TEM. As $J_c$ at low $H$ is similar to the as-grown sample, the estimates of the Larkin length are the same as before and still much smaller than the spacing between clusters. We thus conclude that pinning by the combination of clusters and NPs is also individual rather than collective. Moreover, although the clusters outnumber the NPs by a factor of about 100, their contribution to pinning will be small at low $H$ because they have a smaller $F_P$ and $U_P$ than the NPs, therefore vortices will preferentially occupy the stronger pins. As the vortex density increases and all of the large NP pinning sites become occupied, the pinning by clusters will become progressively more significant. [5,7] On the other hand, a large density of point defects can be seen as a "background" that adds a small contribution to pinning at low $H$ and a large one at high $H$.

Although irradiation-induced defects have a small effect on $J_c$ at low $H$, the clusters and point defects may considerably influence creep in this regime. Ignoring the influence of point defects, the creep process in a system of both NPs and clusters may resemble that in a system consisting only of NPs, yet the comparatively small size of the clusters will result in reduced average pinning energies extracted from creep measurements. That is, both systems will be dominated by strong, single vortex pinning by individual defects, but the system with clusters will have a slightly higher $S$. Point defects will renormalize the energy per unit length of the otherwise unpinned vortex arcs to $\varepsilon_0^{pd} < \varepsilon_0$ to account for the lower condensation energy cost of forming the vortex cores. This will also have the effect of reducing the energy barrier for flux creep. In the limit of $J \to 0$, the reduction will be $\sim 2R_{NP}(\varepsilon_0 - \varepsilon_0^{pd})$. So, the introduction of clusters and point defects will both produce an increase in $S$. For the clusters, this effect will become more pronounced as $H$ increases and more clusters participate in the vortex pinning, while for point defects it will occur at all fields.

We have established that the density of NPs and clusters is not high enough to cause collective pinning in the irradiated sample, however, the question remains as to whether the background of point defects can actuate collective behavior. According to collective creep and vortex-glass theories, [36] $S = (k_B T)/[U_P + \mu k_B T \ln(t/t_o)]$, where $\mu$ is the glassy exponent parameterizing different flux creep regimes, and the Bean critical state forms at a time $t_o$. [41] At high enough temperatures $U_P \ll \mu k_B T \ln(t/t_o)$, which leads to the temperature-independent "plateau" $S \approx [\mu \ln(t/t_o)]^{-1}$; assuming the standard estimate [45,49] for $\ln(t/t_o) \approx 30$, we can approximate $S \approx [30\mu]^{-1}$. The behavior of the irradiated sample is consistent with this possibility; at intermediate temperatures, we extract $\mu(H = 0.3$ T$) \approx$ 1.55 and $\mu(H = 1 \text{ T}) \approx 1.2$, close to the predicted value of $3/2$ for collective creep of small bundles due to weak pinning from uncorrelated disorder caused by random point defects.[41] This agreement with theory suggests that, at intermediate temperatures, the irradiated sample exhibits weak collective pinning.

In the regime of small bundles, the transverse size of the Larkin volume ($R_c$) is larger than $a_0$ and smaller than the penetration depth, $\lambda$, and the longitudinal Larkin length is $L_c^{sb} \approx R_c^2/a_0$. For YBCO at $T \sim 45$ K we have $\lambda \sim 160$ nm, close to the average distance between NPs, indicating that a small bundle could fit *in between* the NPs. In that scenario, it is possible that a fraction of the vortices remain strongly pinned to the NPs, while the creep rates sense the collective dynamics of the remaining vortices. The situation is less clear in the presence of the clusters, which are only $\sim 38$ nm apart. If we assume that $R_c < 38$ nm, then $R_c < a_0$ in the fields under consideration and the conditions for small bundles are not fulfilled. However, it is unclear how the requirements for collective pinning might be different for a system with a range of defects sizes. It is possible that the collective dynamics is robust in the presence of a few clusters inside the Larkin domain. Regarding the as-grown sample, the contribution to pinning from the smaller density of point defects is weak, so the Larkin domain should be very large, containing a large number of NPs and clusters, thus the collective creep regime is never attained.

For the sake of applications, a better understanding of creep in samples with mixed pinning landscapes is necessary because of the relationship between creep and the critical current. As previously mentioned, it is unclear why irradiation is detrimental to $J_c$ at low fields. Fast creep rates at intermediate temperatures may provide a clue as to the reason. $J_c$ reduction in this regime can be readily interpreted as a dynamic effect. Ignoring glassy effects, consider a simple A.K. scenario in which $J_c(t_{obs}) = J_c(t = 0)[1 - S\ln(t_{obs}/t_0)]$. At $T = 45$ K and $\mu_0 H = 0.3$ T, this predicts that $J_c(t_{obs})/J_c(t = 0)$ should be approximately 0.48 for the as-grown sample and 0.35 for the irradiated one. The exact numerical values are not accurate, as we do not have exact knowledge of the evolution of the relaxation process since the formation of the critical state at $t = t_0$. However, the lower predicted value for $J_c(t)/J_c(t = 0)$ in the irradiated sample when considering $S$ suggests that irradiation may have indeed increased $J_c(t = 0)$ at low fields, but by the time we measure it, $J_c$ is lower than the as-grown sample due to faster creep. The drops between $J_c(t=0)$ and $J_c(t_{obs})$ are appreciable, reinforcing our argument for the need to reduce $S$ as an important component of efficacious pinning landscape engineering.

### C. Flux creep after annealing

We now turn to the effects of annealing on the vortex dynamics. Figures 6(d) and (e) show the evolution of $\Delta S$ for all annealing stages. In the as-grown sample, $\Delta S$ decreases very slightly ($S$ increases) upon annealing up to 550°C. The initial decrease at 300°C is likely due to a reduction in oxygen disorder in the strain field [30] around the nanoparticle defects. Changes in the size of the strain field have been seen at 300°C in samples with columnar defects.[20] We measure a slight (7.6% at 0.3 T and 1.6% at 1.0 T) increase in $S$ because the effective size of the pinning site shrinks perhaps due to a reduction in the strain-induced Cooper pair suppression.[50] For $J$ near $J_c$, $F_P \propto \ln(R/\xi_{ab})$.[43] Equating the pinning force with the Lorentz force $J_c B \propto F_P$, $J_c(T_A = 300°C)/J_{c,i} \propto \ln(R_{NP}(T_A = 300°C)/\xi_{ab})/\ln(R_{NP,i}/\xi_{ab})$ where $J_{c,i}$ and $R_{NP,i}$ are the critical current density and NP radius, respectively, prior to annealing. Remarkably, in the as-grown sample, $J_c(T_A = 300°C)/J_{c,i} \approx 0.92$ for a range of conditions ($T$ = 5 K and $T$ = 45 K in fields of 0.3 T, 1.0 T, and 5.0 T), corresponding to a 20% reduction in the nanoparticle radius. Reduced strain around such defects is also thought to result in higher $T_c$,[17] as recorded in Figure 2.

Upon further increasing $T_A$ in the as-grown sample, $\Delta S$ remains constant until $T_A$ = 450 °C, then decreases at 500°C (0.3 T) and 550°C (1.0 T). The last step, annealing the sample at 300°C for 3 hours, fully recovers the original dip depth at 0.3 T and partially (~90%) recovers it at 1.0 T. Reversal of the increase in $S$ during the reoxygenation step suggests that the changes in $S$ due to annealing up to 550°C are predominantly caused by deoxygenation, which results in a 16% and 9% increase in $S$, at 0.3 T and 1 T, respectively. Most importantly, we show that we can indeed reverse the effects of deoxygenation. This will allow us to distinguish the effects of progressive reductions in the irradiation-induced defects from those of deoxygenation when we anneal the irradiated sample.

For the oxygen-irradiated sample, $\Delta S$ increases monotonically with $T_A$ up to 550°C in both fields. At 1.0 T, $\Delta S$ increases in two distinct stages. The first stage corresponds to a reduction of oxygen vacancies around 300°C and should be moderately limited by the small change in NP radius discussed in the case of the as-grown sample. The second corresponds to a reduction of copper vacancies and irradiation-induced defects around 500°C.[19] At 550°C, $\Delta S$ for both the as-grown and irradiated samples are equivalent. The 0.3 T data show a single, broad step; $\Delta S$ is insensitive to $T_A$ up to 350°C, then steadily increases up to 500°C, before slightly decreasing at higher $T_A$. Deoxygenation limits $\Delta S$ due to an increase in the number of oxygen vacancies and is possibly responsible for the decrease in $\Delta S$ at 600°C. The 300°C reoxygenation anneal results in equivalent values for $\Delta S$ in the as-grown and irradiated samples at 0.3 T and similar values at 1.0 T.

In Figure 6(c), we show the 5 T data as it is of interest given the impressive enhancement in $J_c$ at this field. The trend is similar to the low field data; $S$ is increased by irradiation and decreased by annealing. However, $S(T)$ is monotonic in all cases, and the changes in $S$ are small.

Finally, we compare $S(T)$ and $J_c(T)$ for the irradiated sample after different stages of annealing. Revisiting Figure 4 and comparing $J_c(T)$ and $S(T)$ at $T_A$=550°C, we note that $S$ has decreased at the temperatures and fields at which annealing improved $J_c$ ($\mu_0 H$ = 0.3 T, $T \geq 40$ K and $\mu_0 H$ = 1.0 T, $T \geq 65$ K). This coincides with the region where $J_c$ is decreased with irradiation, corroborating the idea that changes in $J_c(t)$ are due to the combined effects of changes in $J_c(t=0)$ and $S$. We further find that annealing allows us to tune the system to a regime in which we see overall improvement in the low-field superconducting properties. After annealing at 450°C, we measure a 10% reduction in the local minimum in $S$ and a concomitant 5% increase in $J_c$(5 K, 0.3 T) and no change in $J_c$(5 K, 1 T), hence preserving the irradiation-induced enhancement.

## IV. SUMMARY AND CONCLUSIONS

In summary, this study investigates the pinning dynamics in an irradiated coated conductor to resolve why remarkable increases in $J_c$ gained by irradiation are frequently accompanied by unfavorable increases in creep. To this end, we first analyze an as-grown sample. We show that the deep and desirable minimum in the creep rate at intermediate temperatures represents a regime of strong individual, rather than collective, pinning and that this strong pinning comes from rare earth oxide nanoparticles. We then study an oxygen-irradiated sample and find that $J_c$ at 5 T is an astonishing 2.5 times higher than in the as-grown sample. Irradiation increases creep at intermediate temperatures such that, instead of a minimum, we measure a high-$S$ plateau implicative of a weak collective or glassy regime. This is striking; irradiation-induced defects essentially reverse the favorable creep properties gained by the nanoparticles, which are present in similar densities in both the as-grown and the irradiated sample. This exemplifies how the introduction of different types of defects that have competing effects on pinning can have adverse effects on creep.

Noting that irradiation introduces a population of few-nm-sized clusters and one of point defects, we elucidate the source of increased creep by performing an annealing study. Annealing allows us to untangle the effects of each type of defect on creep by selectively eliminating point defects. Both clusters and point defects cause increased creep by reducing the activation barrier for vortex depinning from nanoparticles. After discovering the sensitivity in $S(T)$ to the density of point defects and presence of clusters, we determine that clusters are responsible for a modest increase in $S$ within the regime of individual pinning, while the additional point defects result in a stronger increase in $S$ and may in fact induce a transition to the collective regime.

These discoveries have widespread consequences for pinning landscape engineering. First of all, $S(T)$ for the oxygen-irradiated coated conductor studied here is nearly identical to $S(T)$ for a proton- and a gold-irradiated coated conductor. This suggests that the same mechanisms might cause the irradiation-induced increases in creep in all of these systems. Secondly, it is typically thought that HTS materials must be designed specifically for a pre-determined, narrow temperature and field range because the irradiation-induced enhancements will have a limited scope. We suggest that these limitations may be due to the ramifications of the introduction of a high density of weak pinning sites on the creep rate. Concurrently focusing on tuning $S$ as well as $J_c$ may produce more versatile conductors with enhanced properties over a wider range of the $H$-$T$ phase diagram. Thirdly, we demonstrate that our reversible manipulation of the creep rates through a combination of irradiation and annealing adds an extra avenue for pinning landscape engineering. Annealing is a powerful tool that could be used for fundamental purposes, e.g., tuning a crossover from a weak collective to a strong individual pinning regime, or for more applied purposes, e.g., adapting a coated conductor with properties optimized for high-field applications to one appropriate for low-field applications.

## Acknowledgements.

The authors wish to acknowledge illuminating discussions with A. E. Koshelev and I. A. Sadovskyy and useful consultation with A. P. Malozemoff. This work was supported as part of the Center for Emergent Superconductivity, an Energy Frontier Research Center funded

**Figures**

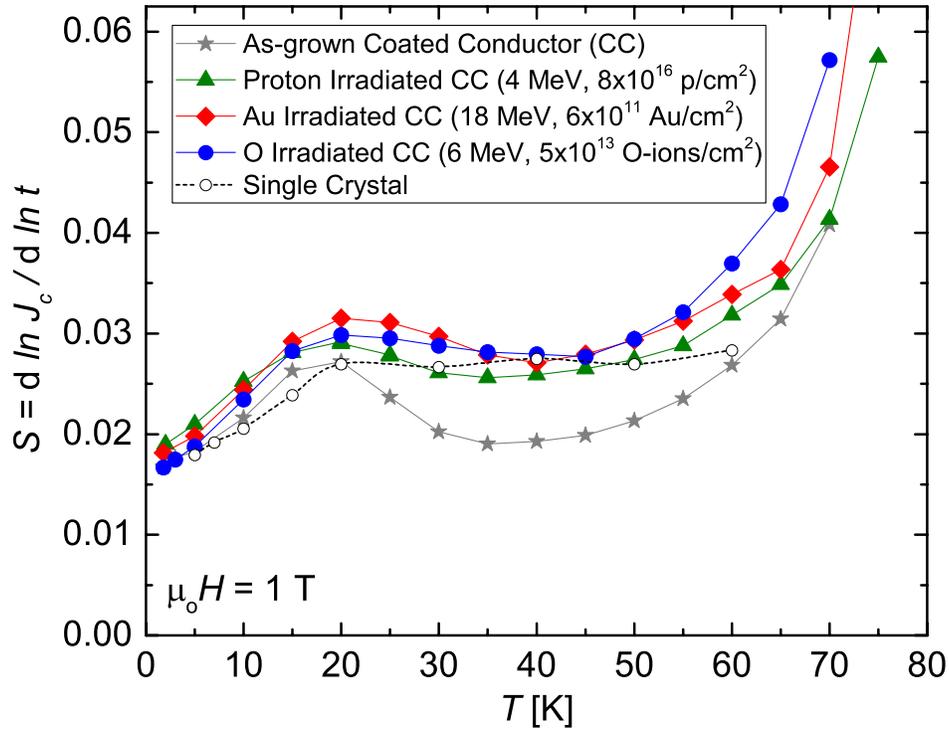

**FIG. 1.** Comparison of temperature-dependent creep rates in as-grown and irradiated (gold, proton, and oxygen) coated conductors prepared by MOD, and a $YBa_2Cu_3O_{7-\delta}$ single crystal (the latter data extracted from Thompson, et al.[2]). The error bars are smaller than the displayed points. All three types of irradiation result in an approximately 50% enhancement in $J_c$ at 1.0 T. The strong pinning effects of nanoparticles create a dip in $S(T)$ at $T$ = 40 K in the as-grown CC. Though the size and density of these nanoparticles is similar in all four CCs, irradiation-induced defects compete with the pre-existing ones, causing unfavorable increases in $S$ at $T$ > 40 K.

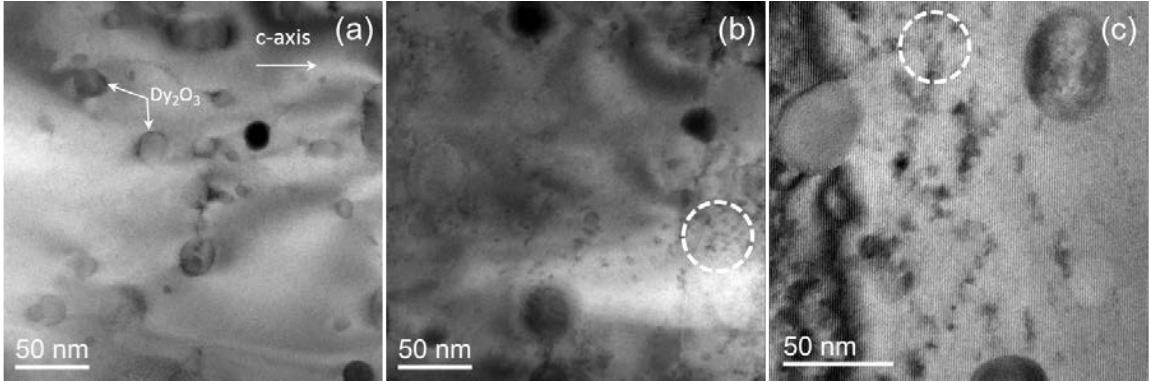

**FIG. 2.** Diffraction contrast transmission electron micrographs for diffraction vector (002) of an (a) as-grown MOD-CC and an (b) oxygen-irradiated MOD-CC similar to the samples presented in this paper. (c) TEM image of the irradiated CC presented here after all annealing stages. Large (20-50 nm diameter) $Dy_2O_3$ nanoparticles (examples are labeled) and stacking faults can be observed in all micrographs. A dispersion of ~5 nm sized clusters are present in micrographs of the pre- and post-annealed O-irradiated CC (the dotted circles encompass examples of regions with many clusters).

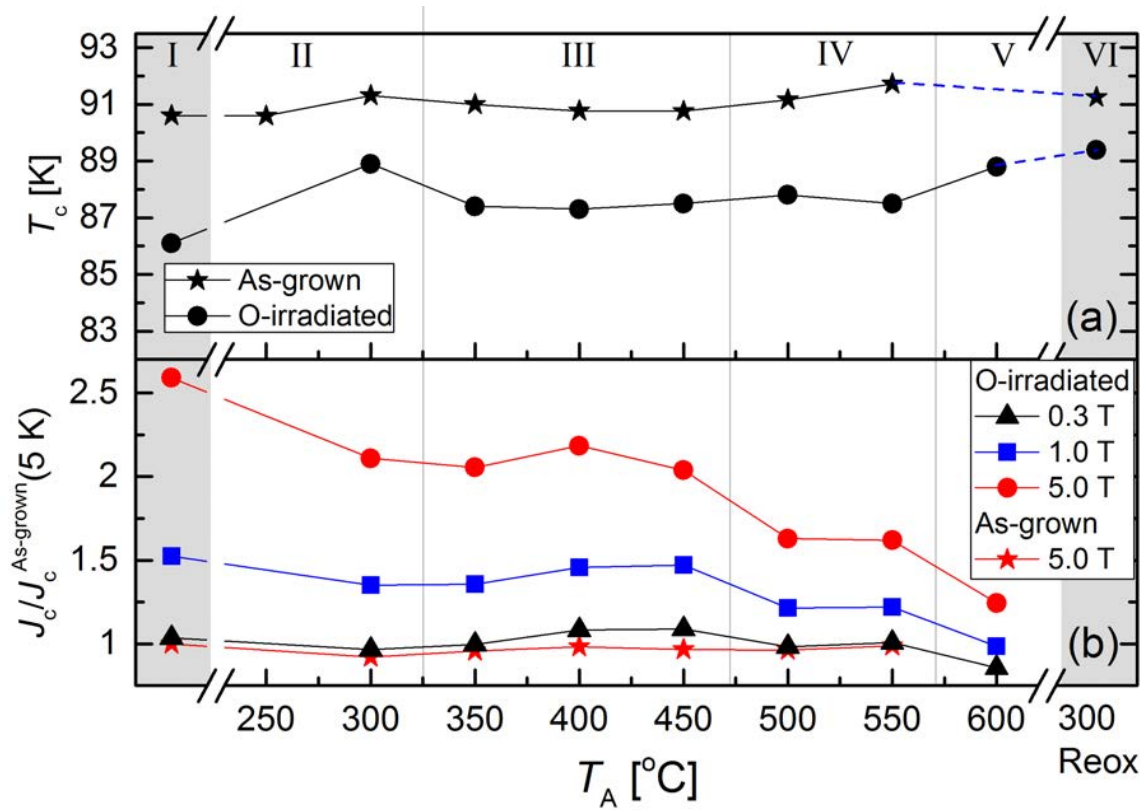

**FIG 3.** (a) Critical temperature ($T_c$) extracted from the midpoint of the temperature dependence of the moment $m(T)$ at $H$ = 2 Oe. (b) Evolution of the critical current ($J_c$) with annealing temperature ($T_A$) at 5 K and $\mu_0 H$ = 0.3 T, 1 T, and 5 T for the oxygen-irradiated sample, and 5.0 T for the as-grown sample, normalized to the pre-annealing value for the as-grown sample. The results for the as-grown sample at 0.3 T and 1.0 T are similar to those at 5.0 T and, for clarity, are not shown.

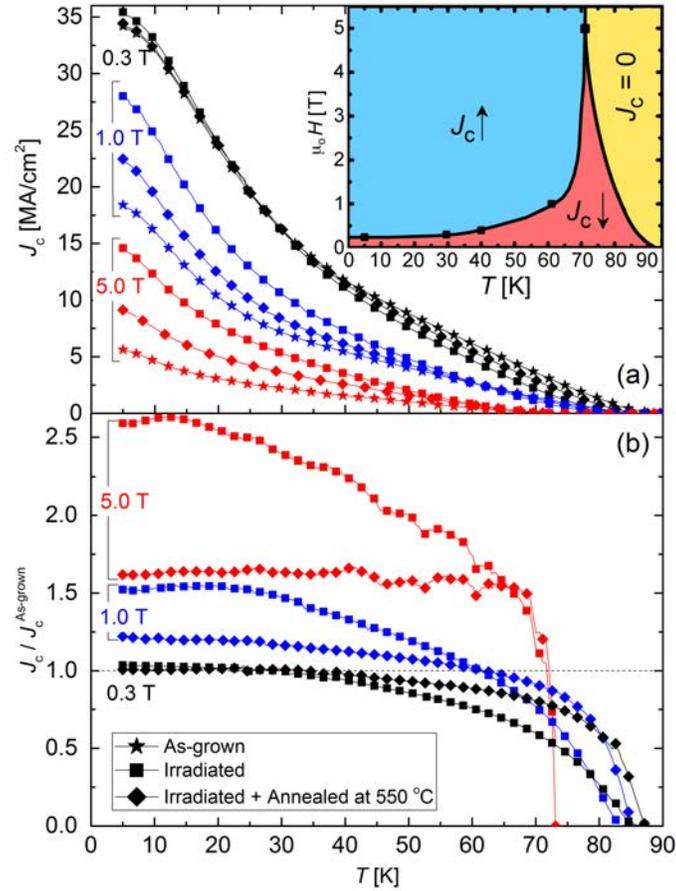

**FIG. 4.** (Upper plot) Temperature dependence of the critical current ($J_c$) at applied magnetic fields of 0.3 T (black curves), 1.0 T (blue), and 5.0 T (red) for the as-grown sample (stars), irradiated sample (squares), and the latter after annealing at 550°C (diamonds). (Lower plot) Ratio between $J_c(T)$ post-irradiation (squares) and post-annealing (diamonds) for the as-grown sample at the indicated fields. (Inset in upper plot) $H$-$T$ Phase diagram showing the regions in which oxygen irradiation is beneficial (blue) vs detrimental (red) to $J_c$, and the bordering non-superconducting region (yellow). The square points were extracted from the data; the solid curves are guides-to-the-eye. The curve separating the blue and red region is $J_c^{Irr}/J_c^{As\text{-}grown}=1$, and the curve separating the red and yellow regions is the irreversibility line.

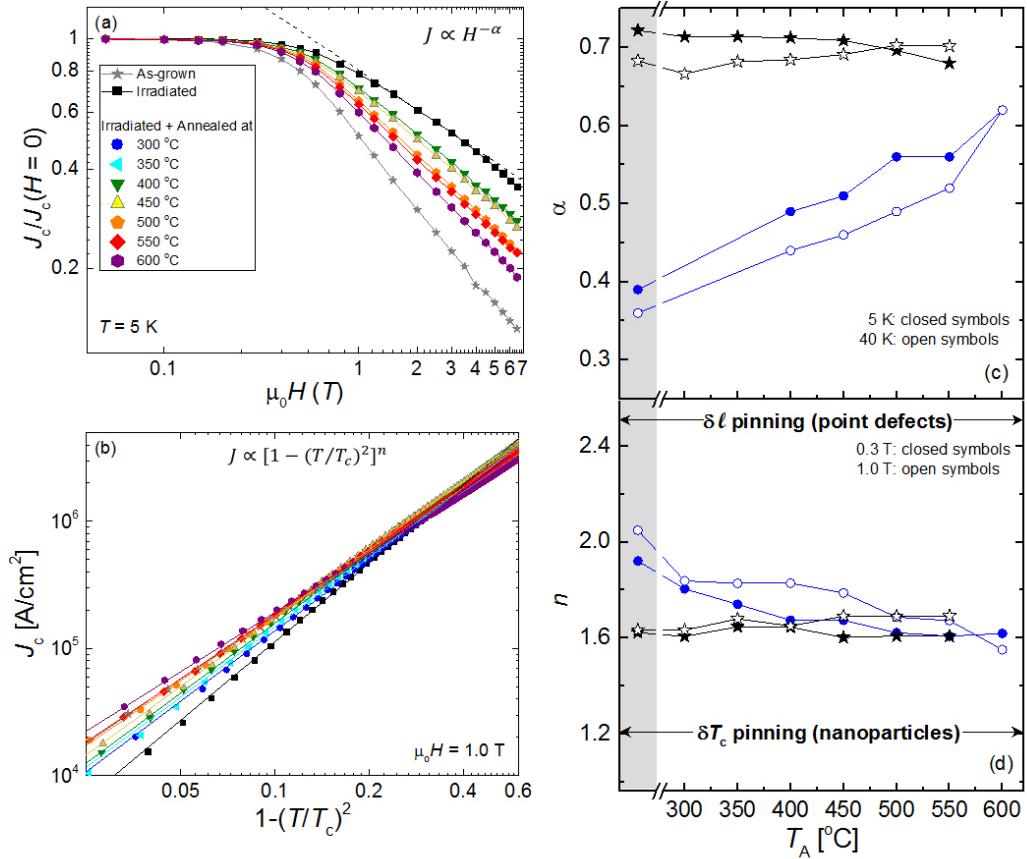

**FIG. 5.** (a) Critical current density ($J_c$) vs magnetic field ($\mu_0 H$) at $T$ = 5 K (upper plot) for an as-grown sample, an oxygen-irradiated sample, and the latter after each annealing step. The dotted blue line shows an example of the region in which $J_c$ exhibits power-law behavior $J_c \propto H^{-\alpha}$. (b) $J_c$ vs $1 - (T/T_c)^2$ for $\mu_0 H$ = 1.0 T on a log-log scale. Solid lines are best linear fits for $T/T_c \geq 0.55$. Evolution of (c) the extracted power law exponent $\alpha$ for $T$=5 K and $T$ = 40 K and (d) $n$ for 0.3 T and 1.0 T with annealing $T$. The gray region highlights the pre-annealing results. The theoretically predicted n-values of 2.5 for $\delta\ell$ pinning (point defects) and 1.2 for $\delta T_c$ pinning (nanoparticles) are indicated in (d).

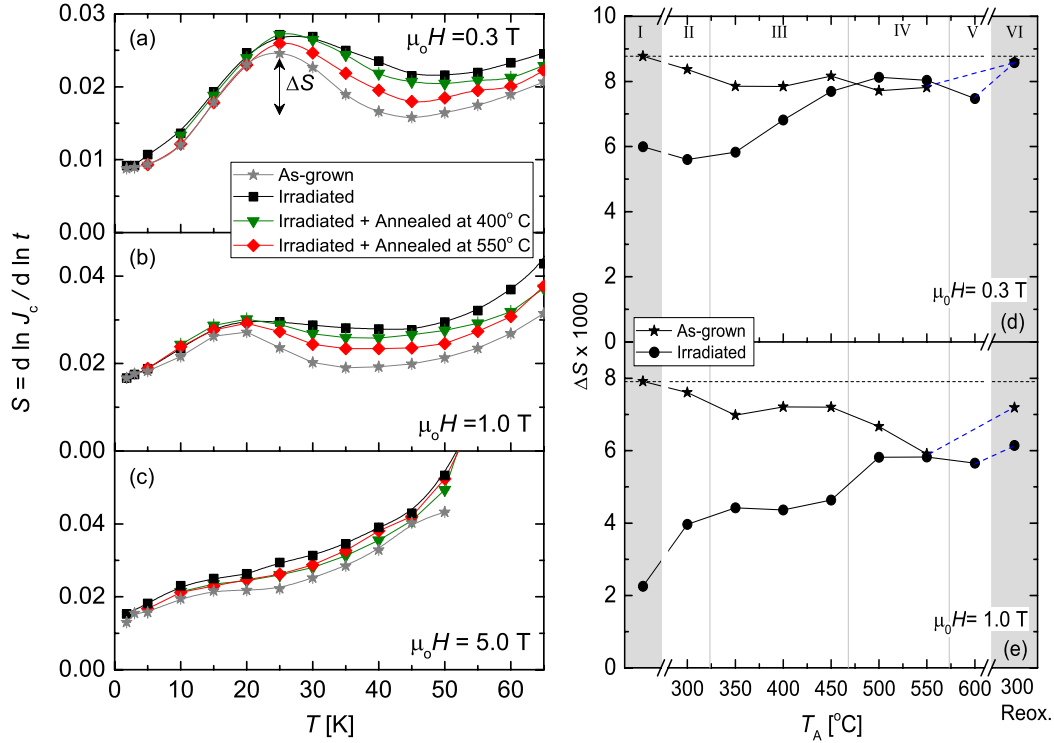

**FIG. 6.** Temperature dependence of the creep rate ($S$) at applied magnetic fields of $\mu_0 H$ = (a) 0.3 T, (b) 1.0 T, and (c) 5.0 T for an as-grown sample, oxygen-irradiated sample, and the latter after each annealing step. Evolution of $\Delta S$ (difference between local maximum and minimum) with increasing annealing temperature $T_A$ for $\mu_0 H$ = (d) 0.3 T and (e) 1.0 T. As an example, the double-headed arrow in (a) marks approximately where the dip depth was measured for the as-grown sample. The gray regions highlight the results before annealing and after the last reoxygenation step. O-irradiation introduces clusters and point defects that result in an increase in $S$ at $T > 20$ K (decrease in $\Delta S$). The point defects are predominantly responsible for this increase in $S$ after irradiation, given that selectively removing point defects through annealing nearly recovers the low $S$ seen in the as-grown sample.

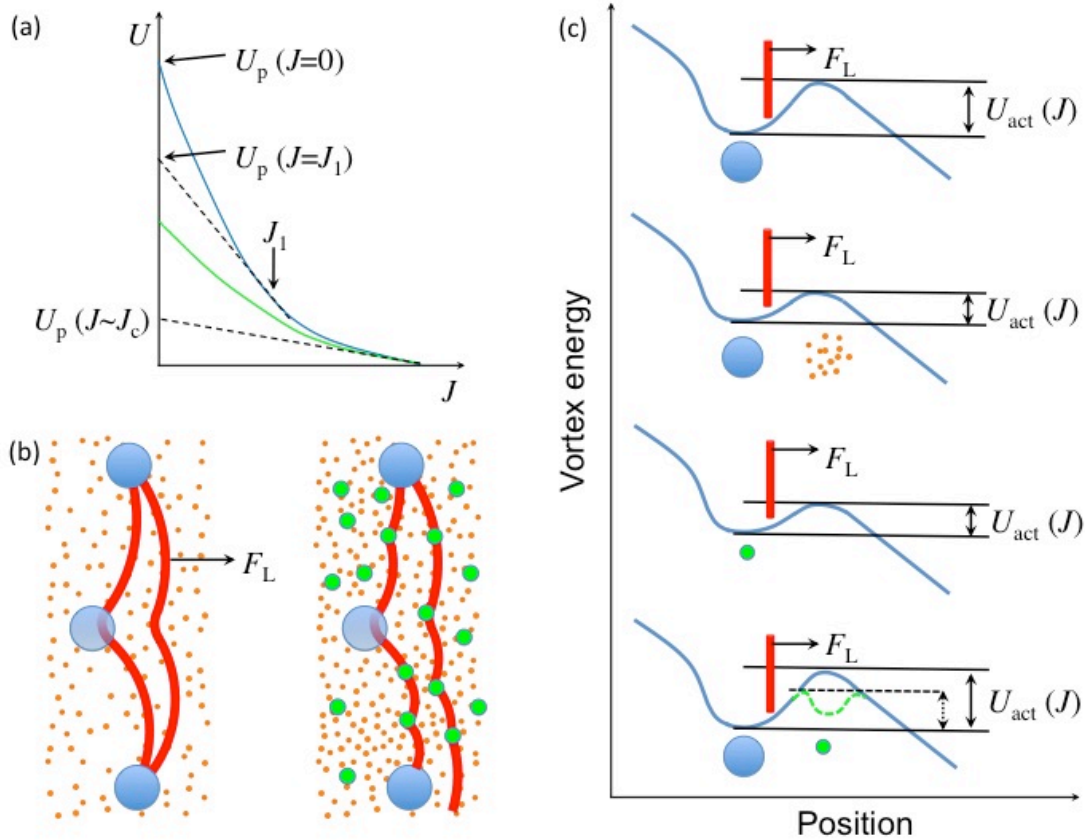

**FIG. 7.** (a) Pinning energy (*U*) vs current density (*J*) for a large nanoparticle precipitate (blue solid line) and a smaller defect such as an irradiation-induced cluster (green solid line). The effective pinning energies for creep at $J\sim J_c$, at an intermediate current $J_1$, and at $J=0$ are indicated. (b) Left panel: Schematic of a possible configuration of a vortex line pinned by a sparse dispersion of large nanoparticle precipitates (large blue circles) and pre-existing point defects (orange circles), and possible changes in configuration due to a current-induced Lorentz force $F_P$. Right panel: Possible changes in the vortex configuration due to admixed clusters (green circles) and additional point defects, both introduced by irradiation. (c) Simplified illustration of the energy barrier for vortex creep. Upper panel: The solid blue curve shows the barrier when only nanoparticles are considered. Second from top: Lowering of the energy barrier due to a background of point defects. The lower two panels illustrate two possible ways in which additional clusters could result in a lower barrier to creep than only nanoparticles; (second from bottom) a vortex could depin from a cluster, which has lower pinning energy than a nanoparticle, and (bottom panel) a cluster could facilitate depinning from a larger nanoparticle.